\documentclass{svproc}
\usepackage{amsfonts,amssymb}
\usepackage{xspace}
\usepackage{graphicx}
\usepackage{amsmath}

\newcommand{\GeV}{{\ensuremath\rm GeV}\xspace}
\newcommand{\TeV}{{\ensuremath\rm TeV}\xspace}
\newcommand{\lb}{\left(}
\newcommand{\rb}{\right)}
\newcommand{\al}{\alpha}

\newcommand{\fb}{{\ensuremath\rm fb}\xspace}
\newcommand{\HS}{\texttt{HiggsSignals}}
\newcommand{\HB}{\texttt{HiggsBounds}}

\newcommand{\eqn}{equation}
\usepackage{url}

\begin{document}
\mainmatter              
\title{Constraining extended scalar sectors
at current and future colliders - an update}
\titlerunning{Extended scalars update}  
%
\author{Tania Robens\inst{1}}
\authorrunning{Tania Robens} 
%
%
\institute{Rudjer Boskovic Institute, Bijenicka cesta 54, 10000 Zagreb, Croatia\\
\email{trobens@irb.hr}}

\maketitle              

\begin{abstract}
In this proceeding, I discuss several models that extend the scalar sector of the Standard Model by additional matter states. I here focus on results for models with singlet extensions, which have been obtained recently and update some of the results presented in previous work. In more detail, I will briefly review the option to test a strong first-order electroweak phase transition using precision measurements in the electroweak sector, as well as production cross-sections for non-standard scalar production at Higgs factories.\\
RBI-ThPhys-2022-36
\keywords{new physics models, extended scalar sectors, future facilities}
\end{abstract}

\section{Introduction}
In this proceeding, I discuss various new physics models that extend the Standard Model (SM) by adding additional fields that transform as singlets under the SM gauge group. The proceedings contain results which can be seen as follow-ups of the studies presented e.g. in \cite{Robens:2022zav}. I will therefore focus on novel developments within the last year. In particular, I discuss
\begin{itemize}
\item{}The real singlet extension of the SM, which comes with an additional scalar that transforms as a singlet. The model features one additional CP even neutral scalar. See \cite{Pruna:2013bma,Robens:2015gla,Robens:2016xkb,Ilnicka:2018def,DiMicco:2019ngk} for original literature as well as \cite{Robens:2022oue,Robens:2022piz} for results presented here;
\item{}The two real singlet extension (TRSM), where the SM scalar sector is extended by two additional gauge singlets, featuring in total three CP even neutral scalars that also allow for interesting cascade decays. Original literature can be found in \cite{Robens:2019kga}, while the results discussed here have first been presented in \cite{Robens:2022nnw}.
\end{itemize}

All models are confronted with most recent theoretical and experimental constraints. Theory constraints include the minimization of the vacuum as well as the requirement of vacuum stability and positivity. We also apply constraints from perturbative unitarity and perturbativity of the couplings at the electroweak scale.

Experimental bounds include the agreement with current measurements of the properties of the 125 \GeV~ resonance discovered by the LHC experiments, as well as agreement with the null-results from searches for additional particles at current or past colliders. We additionally impose constraints from electroweak precision observables (via $S,\,T,\,U$  parameters \cite{Altarelli:1990zd,Peskin:1990zt,Peskin:1991sw}). For our studies, we use a combination of private and public tools. In particular, we use \HB~ \cite{Bechtle:2008jh,Bechtle:2011sb,Bechtle:2013wla,Bechtle:2020pkv}  and \HS~ \cite{Bechtle:2013xfa,Bechtle:2020uwn} for the comparison with current collider constraints. Experimental numbers are taken from \cite{Baak:2014ora,Haller:2018nnx} for electroweak precision observables. Some predictions for production cross sections shown here have been obtained using Madgraph5 \cite{Alwall:2011uj}.

\section{Real singlet extension}
As a first simple example, we discuss a real singlet extension of the SM with a $\mathbb{Z}_2$ symmetry previously reported on in \cite{Pruna:2013bma,Robens:2015gla,Robens:2016xkb,deFlorian:2016spz,Ilnicka:2018def,DiMicco:2019ngk}. The $\mathbb{Z}_2$ symmetry is softly broken by a vacuum expectation value (vev) of the singlet field, inducing mixing between the gauge-eigenstates which introduces a mixing angle $\al$. The model has in total 5 free parameters. Two of these are fixed by the measurement of the $125\,\GeV$ resonance mass and electroweak precision observables. The free parameters of the model are then given by
\begin{\eqn}
\sin\al,\,m_2,\,\tan\beta\,\equiv\,\frac{v}{v_s},
\end{\eqn}
with $v\,(v_s)$ denoting the doublet and singlet vevs, respectively. We concentrate on the case where $m_2\,\geq\,125\,\GeV$, where SM decoupling corresponds to $\sin\al\,\rightarrow\,0$.

\subsection{Current constraints}
\begin{center}
\begin{figure}
\begin{center}
\begin{minipage}{0.48\textwidth}
\includegraphics[width=\textwidth]{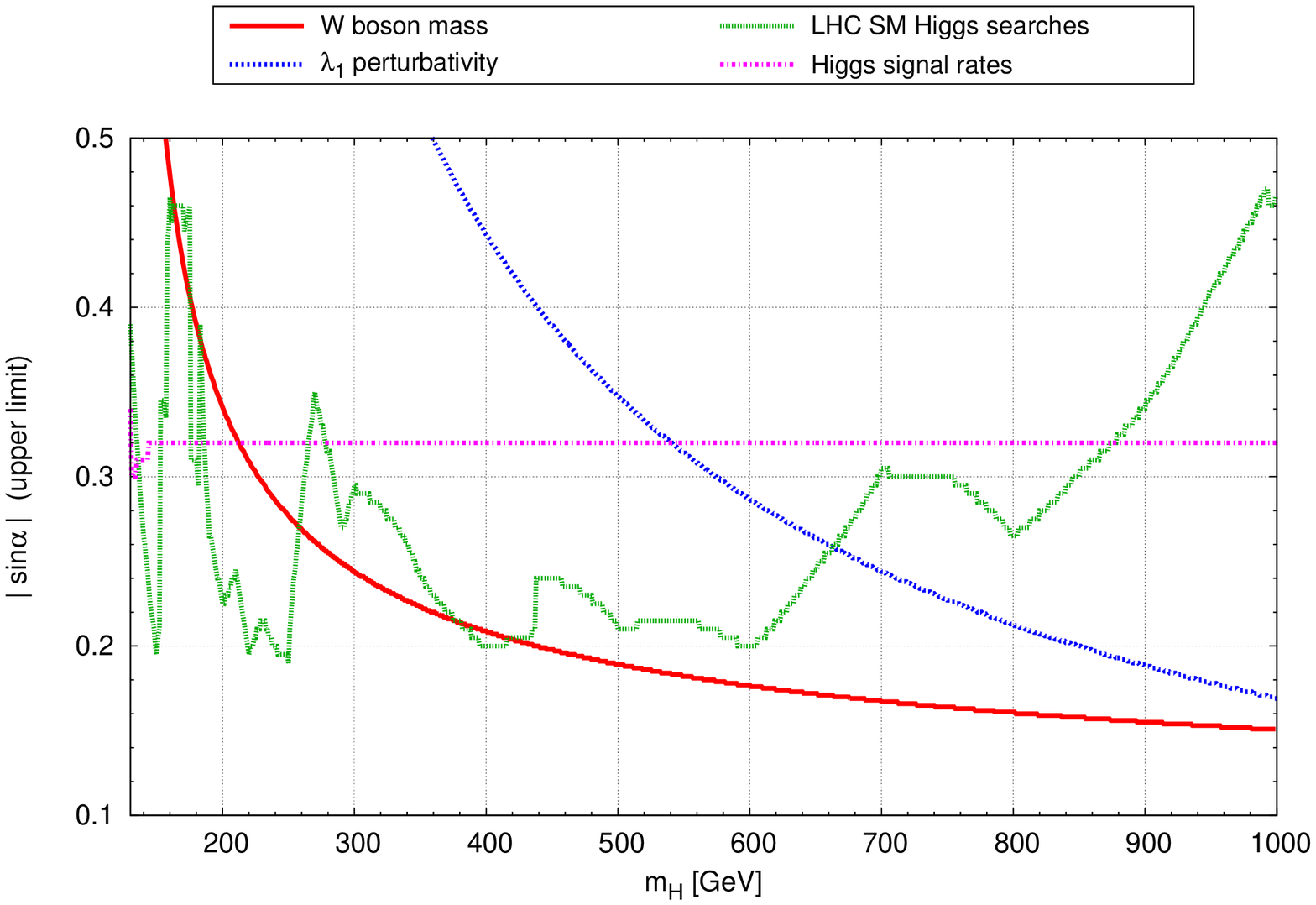}
\end{minipage}
\begin{minipage}{0.42\textwidth}
\includegraphics[width=\textwidth]{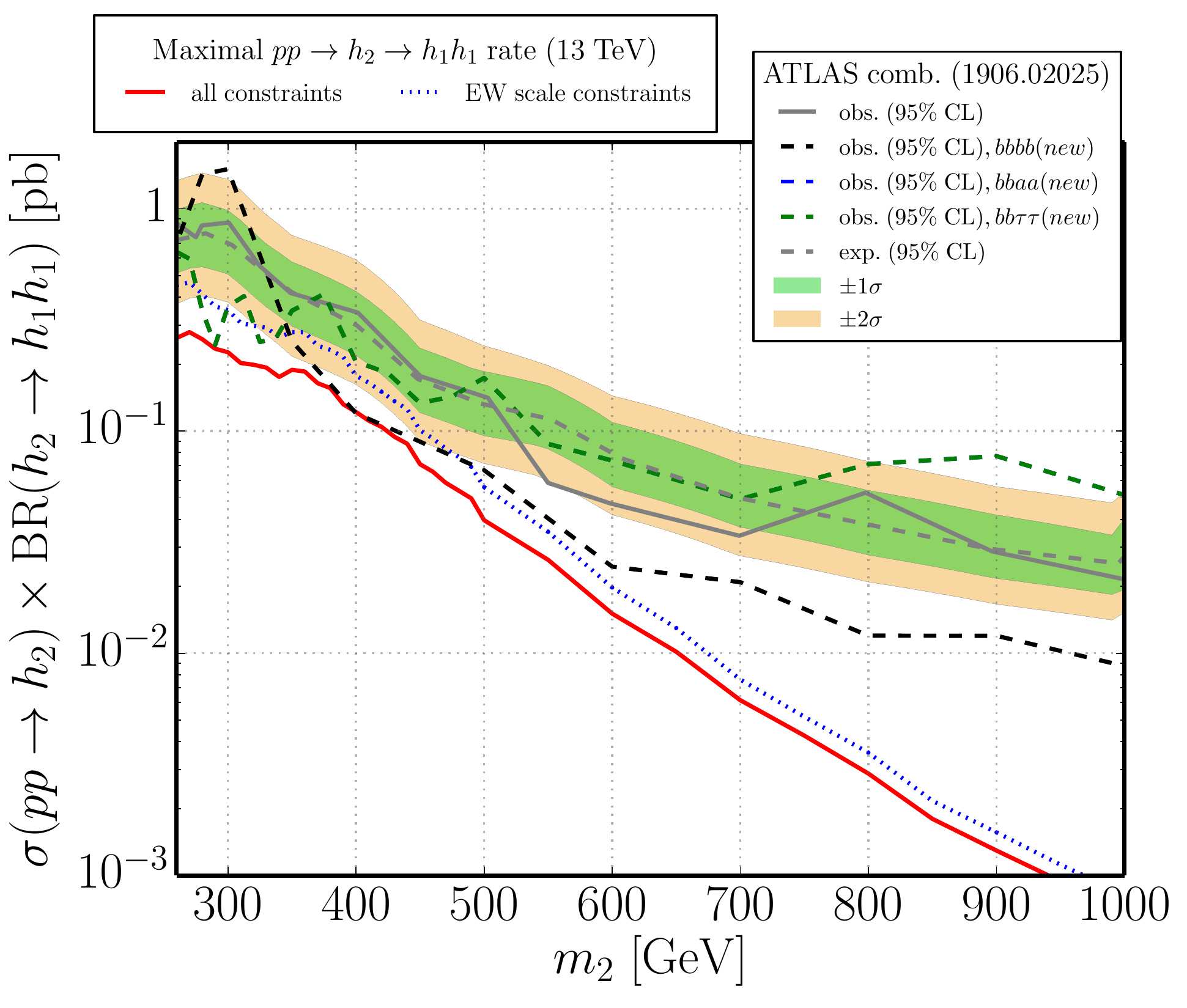}
\end{minipage}
\caption{Results for the singlet extension, {\sl Left:} comparison of current constraints for a fixed value of $\tan\beta\,=\,0.1$, taken from \cite{Robens:2022oue}. {\sl Right:} maximal $H\,\rightarrow\,h\,h$ allowed, with electroweak constraints at the electroweak scale (blue) or including RGE running to a higher scale (red), in comparison with results from the ATLAS combination and direct searches. Taken from \cite{Robens:2022piz}.}\label{fig:singlet}
\end{center}
\end{figure}
\end{center}

Limits on this model are shown in figure \ref{fig:singlet}, including a comparison of the currently maximal available rate of $H\,\rightarrow\,h_{125}h_{125}$ with the combination limits from ATLAS \cite{Aad:2019uzh} as well as dedicated searches with full run II data \cite{ATLAS:2022hwc,ATLAS:2021ifb,ATLAS-CONF-2021-030}\footnote{I thank D. Azevedo for providing exclusion limits for $b\,\bar{b}\tau^+\tau^-$ \cite{ATLAS-CONF-2021-030} in a digitized format, as used in \cite{Abouabid:2021yvw}.}. The most constraining direct search bounds are in general dominated by searches for diboson final states \cite{CMS-PAS-HIG-13-003,Khachatryan:2015cwa,Sirunyan:2018qlb,Aaboud:2018bun}. In some regions, the Run 1 Higgs combination \cite{CMS-PAS-HIG-12-045} is also important. Especially \cite{Sirunyan:2018qlb,Aaboud:2018bun} currently correspond to the best probes of the models parameter space\footnote{We include searches currently available via \HB.}.

\subsection{Strong first-order electroweak phase transition}
One question currently of high interest is whether new physics will enable a strong first-order electroweak phase transition. In general, this is studied in the context of exotic Higgs decays, see e.g. \cite{Carena:2022yvx} for recent work within the Snowmass process. However, one can also test this in models where the second scalar is heavier than the 125 \GeV resonance, as e.g. discussed in \cite{Papaefstathiou:2020iag,Papaefstathiou:2021glr}. 

In \cite{Papaefstathiou:2022oyi}, we have investigated whether regions in the models parameter space that allow for a strong first-order electroweak phase transition can be tested using electroweak precision measurements, in particular, the Higgs signal strength as well W-boson mass, where we resort to the PDG value \cite{ParticleDataGroup:2020ssz} $m_W^\text{exp}\,=\,80.379\,\pm\,0.012\,\GeV$\footnote{This corresponds to the PDG value at the time of the above reference. The current value \cite{Workman:2022ynf} is slightly lower. We do not expect this to have a qualitatively large impact.}. In our study the $\mathbb{Z}_2$ symmetry is not imposed, leading to additional terms in the potential with respect to the singlet scenario discussed above.

\begin{center}
\begin{figure}[th!]
\begin{center}
\includegraphics[width=0.8\textwidth]{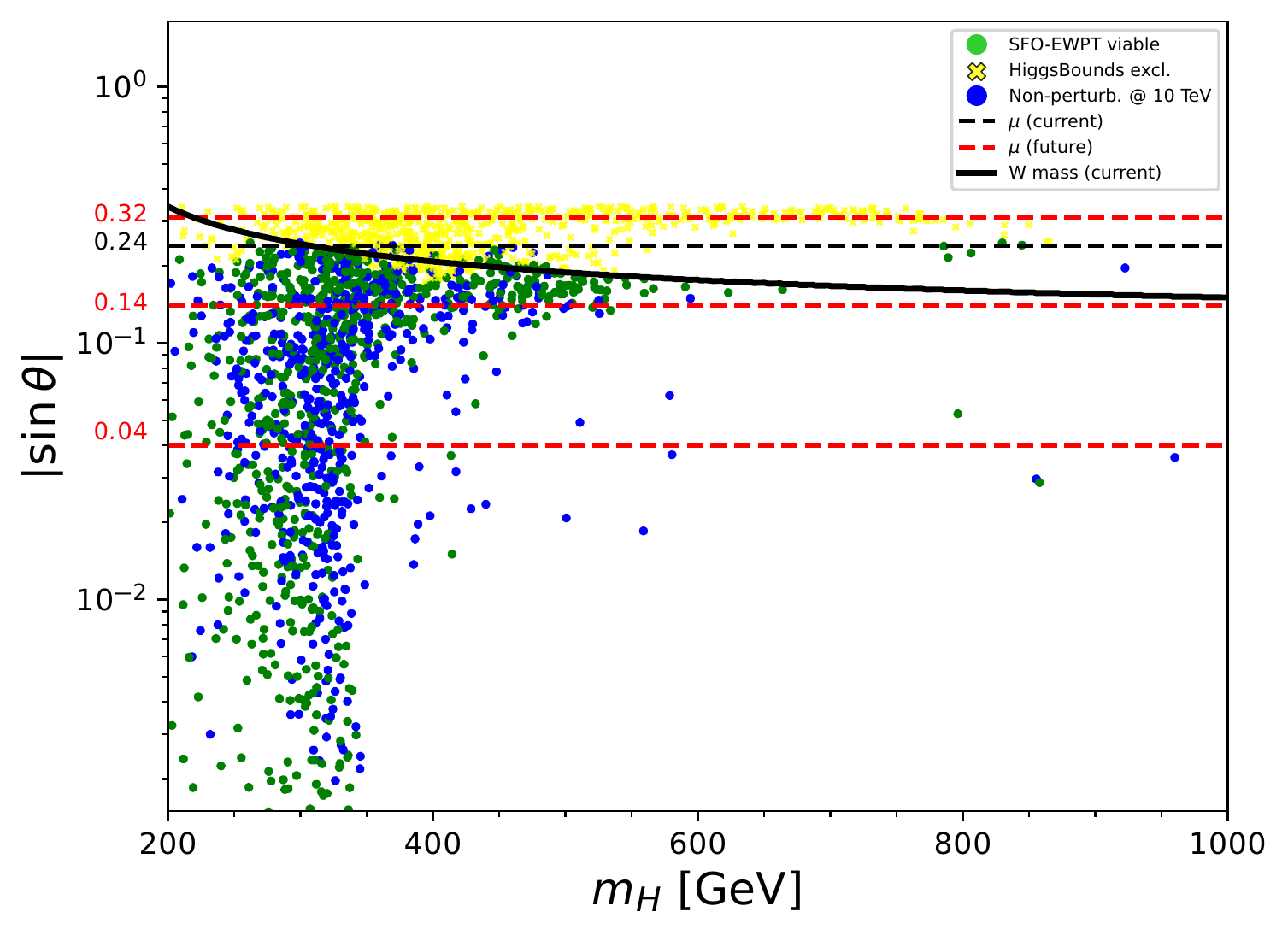}
\caption{\label{fig:} The viable points for a SFO-EWPT are shown in
  lime green filled circles. The points denoted by yellow crosses yield the necessary conditions for a SFO-EWPT, but are excluded
by direct searches for heavy scalars (imposed via the
\texttt{HiggsBounds} package). The points denoted by blue circles are
those allowed by direct searches for heavy scalars but that become non-perturbative at 10~TeV. The current W-boson mass constraint is
shown in solid black and the constraints due to signal strength
measurements of the SM-like Higgs are shown in dashed lines, with
black indicating the current constraint at 95\% C.L. ($\left| \sin \theta \right| =
0.24$) and red indicating the corresponding future constraints assuming $5\%,\,1\%,$ and
$0.1\%$ measurements with a central value of  $\mu\,=\,1$ ($\left| \sin \theta \right| =
0.32, 0.14, 0.04$, respectively). Figure taken from \cite{Papaefstathiou:2022oyi}. }
\end{center}
\end{figure}
\end{center}

In general, constraints depend not only on the mass and mixing angle, but also additional couplings in the potential that govern scalar self-interactions. However, two constraints that only depend on the second scalar mass and the mixing angle\footnote{At the order of perturbation theory discussed here; extending to higher orders might introduce additional parameter dependencies.} are the one-loop corrections to the W-boson mass, as well as the signal strength measurements for the 125~\GeV{} scalar. The latter is related to the mixing angle via $\cos^2\theta\,\gtrsim\,\mu$. In \cite{Papaefstathiou:2022oyi}, we compared to the by that time current ATLAS combination value \cite{ATLAS-CONF-2021-053}, $\mu\,=\,1.06\,\pm\,0.06$, leading to $|\sin\theta|\,\lesssim\,0.24$ at 95\% C.L.. \footnote{Note that the Run 2 combinations of ATLAS \cite{ATLAS:2022vkf} and CMS \cite{CMS:2022dwd} separately lead to $|\sin\theta|\,\lesssim\,0.26$  and $|\sin\theta|\,\lesssim\,0.33$, respectively.} We display this bound as ``current bound" in figure \ref{fig:}. Furthermore, the precision that might be achievable at future colliders for the signal strength varies from collider to collider and can reach per-mille level at future machines \cite{caterinatalk,Dawson:2022zbb}. This is shown in lines corresponding to 95\% C.L., for an assumed precision of $5\%,\,1\%,$ and $0.1\%$ in figure \ref{fig:}. For these, set $\mu\,=\,1$. 
For direct searches, we only impose \textit{current} constraints through the \texttt{HiggsBounds} package. 

For the calculation of the contributions to the W-boson mass, we essentially follow the work presented in \cite{Lopez-Val:2014jva}. However, since then many of the parameters used in the evaluation of the SM-like contribution have been updated, we have re-evaluated the SM-prediction~\cite{Awramik:2003rn} using the most recent electro-weak parameters~\cite{Keshavarzi:2019abf,ParticleDataGroup:2020ssz}, leading to the theoretical prediction \cite{Papaefstathiou:2022oyi} $m_W^\text{SM}\,=\,80.356\,\GeV$. 

We see that even for permill-precision of the coupling strength, viable parameter points in this model remain which can lead to a strong first-order electroweak phase transition.

\section{Two real singlet extension}

I now turn to a model where the scalar sector of the SM has been augmented by two real scalar fields, which obey a $\mathbb{Z}_2\,\otimes\,\mathbb{Z}_2'$ symmetry. This model has first been presented in \cite{Robens:2019kga}, with recent reviews in \cite{Robens:2022lkn,Robens:2022nnw}.

While previous studies were concentrating on physics at hadron colliders, I now want to report on the options to investigate the model at future Higgs factories, along the lines of models discussed in \cite{Robens:2022zgk}. In particular, I enhance the benchmark planes presented in the above references to a general scan and show production cross sections achievable at an $e^+e^-$ collider with a center-of-mass (com) energy of $250\,\GeV$\footnote{The parameter scans include only current bounds, not possible discovery or exclusion at e.g. a HL-LHC.}.

The particle content of this model in the scalar sector consists of 3 CP-even neutral scalars, where mass eigenstates correspond to admixtures of the gauge eigenstates.  In the following, we will use the following mass hierarchy
\begin{\eqn}\label{eq:hier}
M_1\,\leq\,M_2\,\leq\,M_3
\end{\eqn}
and denote the corresponding physical mass eigenstates by $h_i$.
Gauge and mass eigenstates are related via a mixing matrix. The model contains in total 9 free parameters, out of which 2 are fixed by the observation of a scalar particle with the mass of 125 \GeV~ as well as electroweak precision observables. Apart from the masses, also the vaccum expectation values and mixing angles serve as input parameters. We scan the parameter space using the implementation into \texttt{ScannerS} \cite{Coimbra:2013qq,Muhlleitner:2020wwk}.
\subsection{Production of light scalars at Higgs factories}
\begin{center}
\begin{figure} 
\begin{center}
\includegraphics[width=0.45\textwidth]{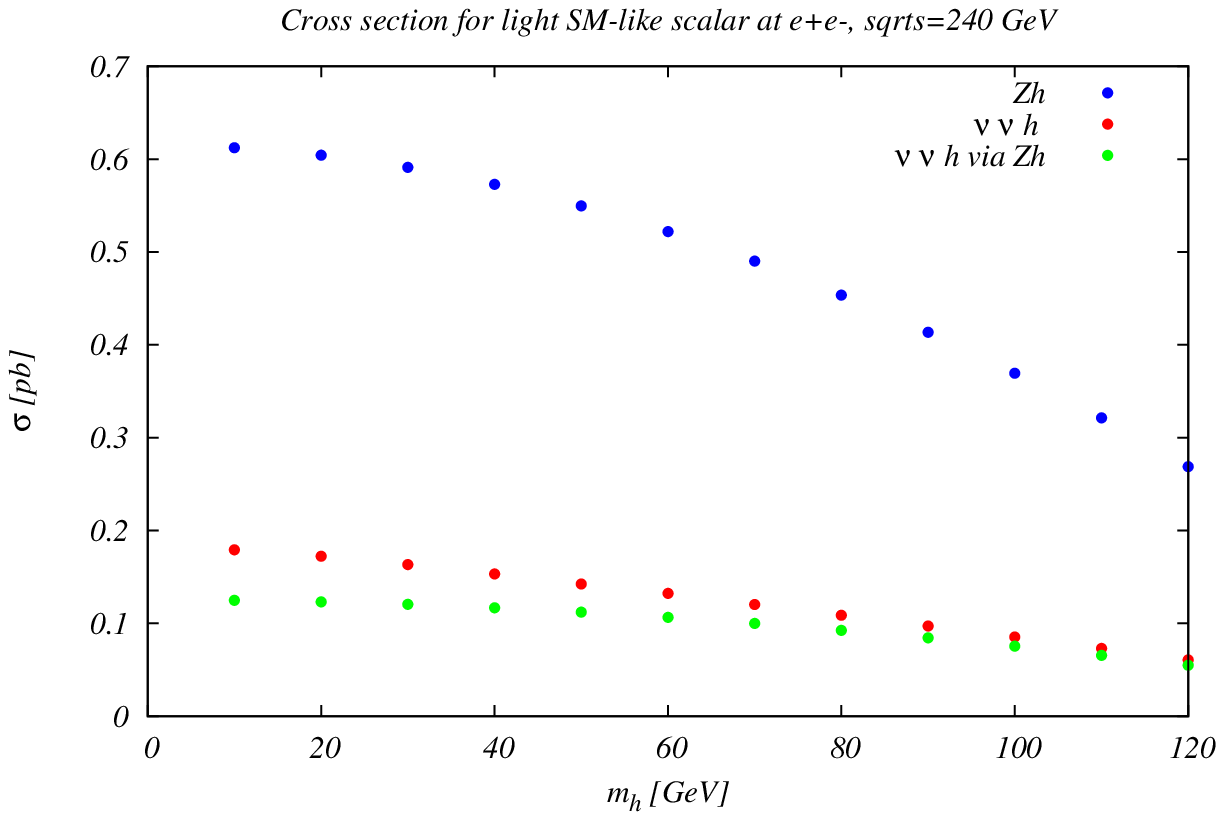}
\includegraphics[width=0.45\textwidth]{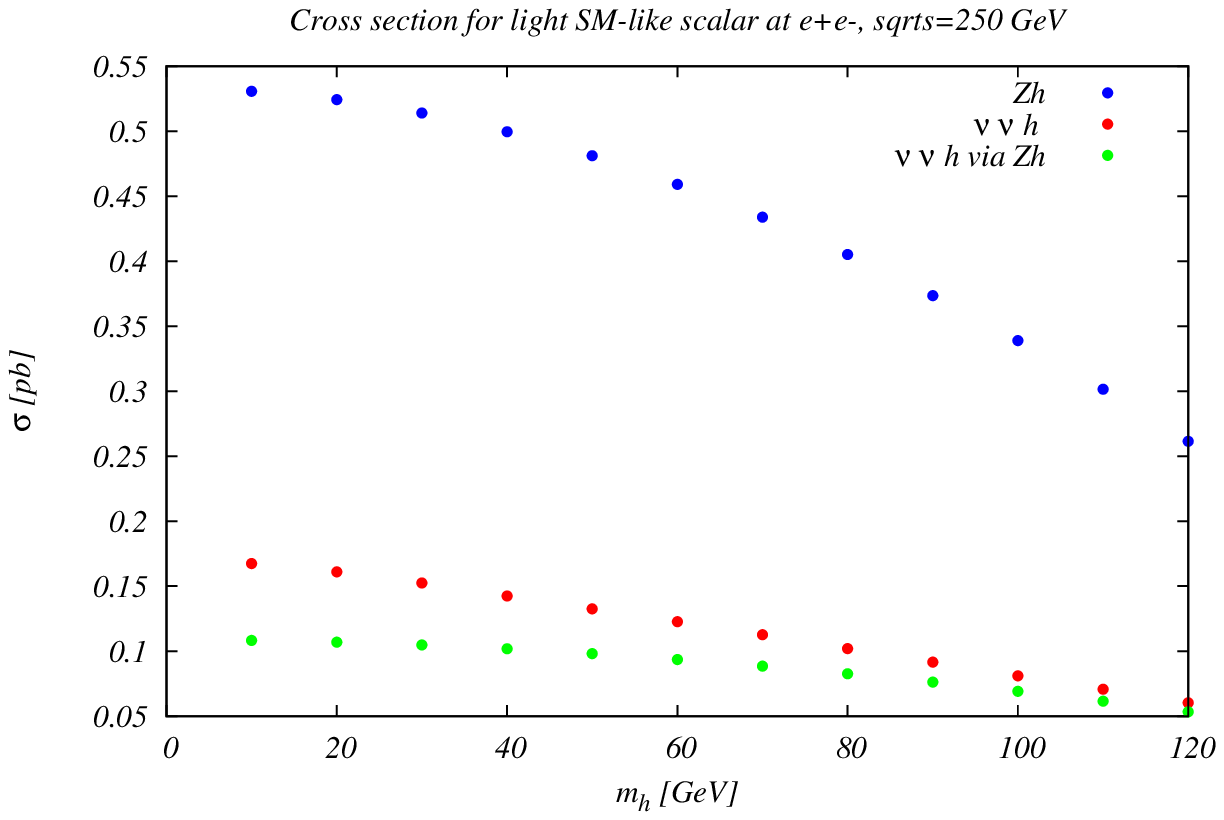}
\caption{\label{fig:prod250} Leading order production cross sections for $Z\,h$ and $h\,\nu_\ell\,\bar{\nu}_\ell$ production at an $e^+\,e^-$ collider with a com energy of 240 \GeV {\sl (left)} and 250 \GeV~ {\sl (right)} using Madgraph5 for an SM-like scalar h. Shown is also the contribution of $Z\,h$ to $\nu_\ell\,\bar{\nu}_\ell\,h$ using a factorized approach for the Z decay. Taken from \cite{Robens:2022zgk}.}
\end{center}
\end{figure}
\end{center}

In this model, the only feasible production is $Zh$ radiation of the lighter scalar, with production cross sections given in figure \ref{fig:prod250}. Cross sections have been derived using Madgraph5 \cite{Alwall:2011uj}.

\begin{center}
\begin{figure} 
\begin{center}
\includegraphics[width=0.45\textwidth]{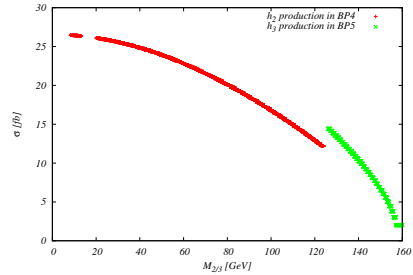}
\end{center}
\caption{\label{fig:prod45} Production cross sections for $Z h_{2/3}$ in BPs 4 and 5, respectively, at a 250 \GeV~ Higgs factory. Taken from \cite{Robens:2022nnw}.}
\end{figure}
\end{center}

We can now investigate what would be production cross sections for scalar particles with masses $\lesssim\,160\,\GeV$ at Higgs factories.
Rates for the production of a 125 \GeV scalar with subsequent decays into novel scalar states have been presented in \cite{Robens:2022lkn,Robens:2022nnw} and will not be repeated here. Instead, we turn to the Higgs-Strahlung production of new physics scalars, where we first present results for the benchmark planes presented in \cite{Robens:2019kga}. If we require production rates of $Z\,h_i$ to be larger than $\sim\,10\,\fb$, only BPs 4 and 5 render sufficiently large rates for the production of $h_2$ and $h_3$, respectively. Production rates are independent of the other scalars, and we therefore depict them for both BPs in figure \ref{fig:prod45}.
BP4 is constructed in such a way that as soon as the corresponding parameter space opens up, the $h_1\,h_1$ decay becomes dominant; final states are therefore mainly $Z\,b\bar{b}b\bar{b}$ if $M_2\,\gtrsim\,2\,M_1$. Below that threshold, dominant decays are into a $b\,\bar{b}$ pair, which means that standard searches as e.g. presented in \cite{Drechsel:2018mgd,Wang:2020lkq} should be able to cover the parameter space.

Similarly, in BP5 the $h_3\,\rightarrow\,h_1\,h_1$ decay is also favoured as soon as it is kinematically allowed. Therefore, in this parameter space again $Z b\bar{b}b\bar{b}$ final states become dominant. Otherwise $Z\,b\bar{b}$ and $ZW^+W^-$ final states prevail, with a cross over for the respective final states at around $M_3\,\sim\,135\,\GeV$. Branching ratios for these final states are in the $40-50\%$ regime.

\subsection{More general scan}

\begin{center}
\begin{figure}
\includegraphics[width=0.45\textwidth]{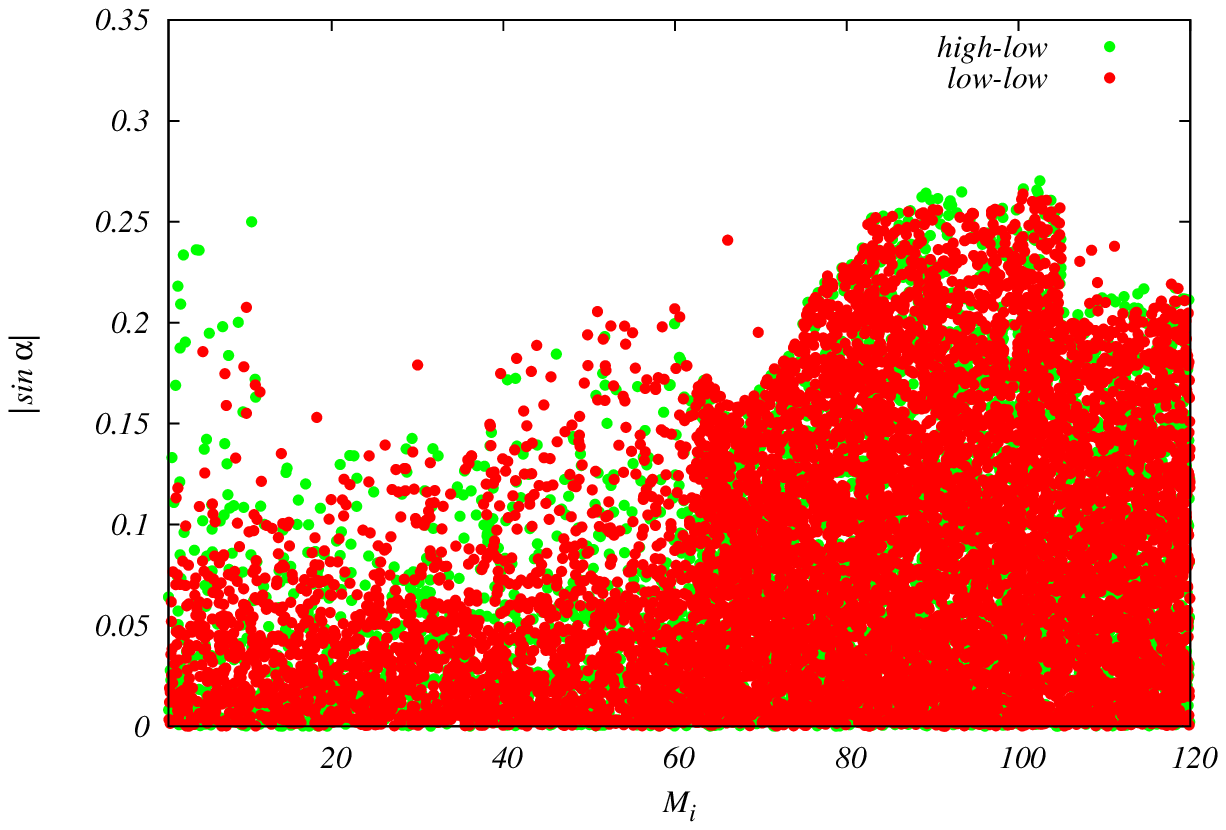}
\includegraphics[width=0.45\textwidth]{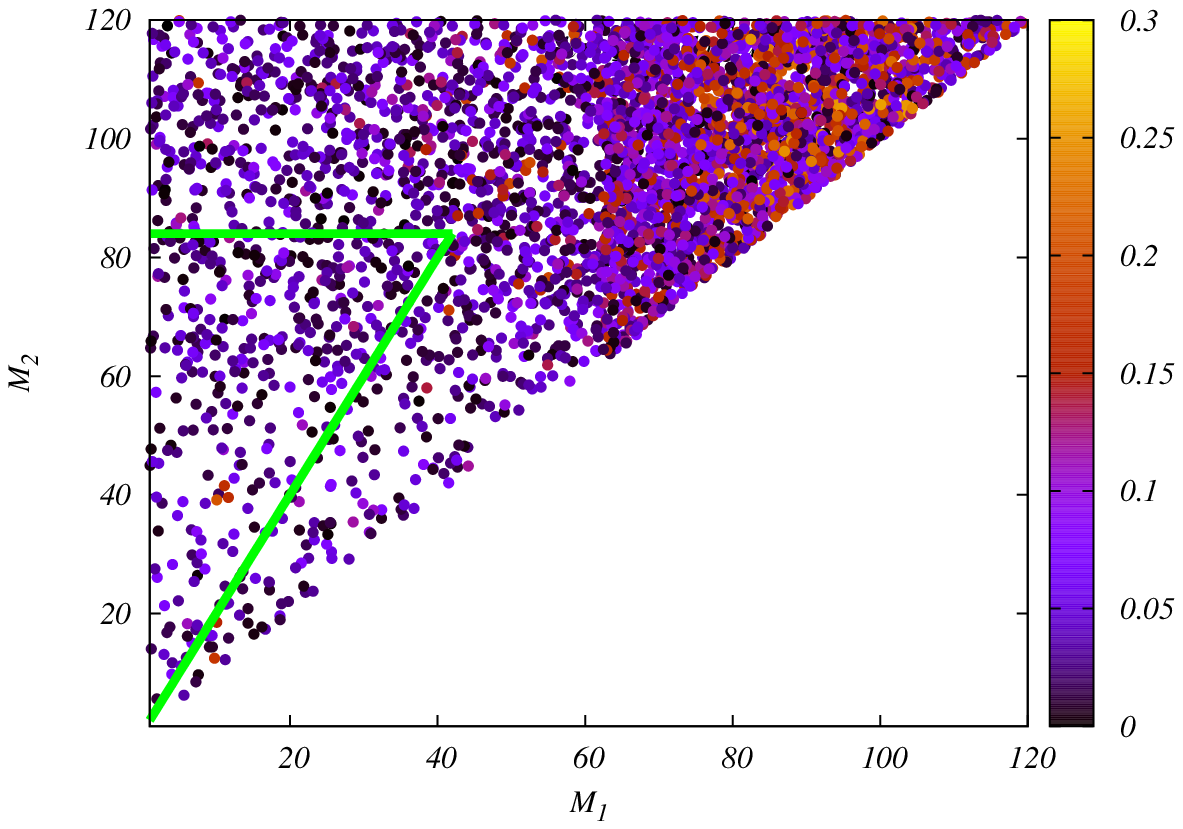}
\caption{\label{fig:trsm} Available parameter space in the TRSM, with one (high-low) or two (low-low) masses lighter than 125 \GeV. {\sl Left}: light scalar mass and mixing angle, with $\sin\al\,=\,0$ corresponding to complete decoupling. {\sl Right:} available parameter space in the $\lb m_{h_1},\,m_{h_2}\rb$ plane, with color coding denoting the rescaling parameter $\sin\al$ for the lighter scalar $h_1$. Within the green triangle, $h_{125}\,\rightarrow\,h_2 h_1\,\rightarrow\,h_1\,h_1\,h_1$ decays are kinematically allowed. Taken from \cite{Robens:2022zgk}.}
\end{figure}
\end{center}

In the following, I present results for a more general parameter scan of the masses, taken from \cite{Robens:2022erq,Robens:2022zgk}, given in figure \ref{fig:trsm}. In this figure, two data-sets are considered which fulfill all current constraints as implemented using the current versions of \texttt{ScannerS} and \HB, \HS. The are labelled "low-low" if both $M_{1,2}\,\leq\,125\,\GeV$ and "high-low" if $M_1\,\leq\,125\,\GeV,\,M_3\,\geq\,125\,\GeV$.

\begin{center}
\begin{figure}
\begin{center}
\includegraphics[width=0.49\textwidth]{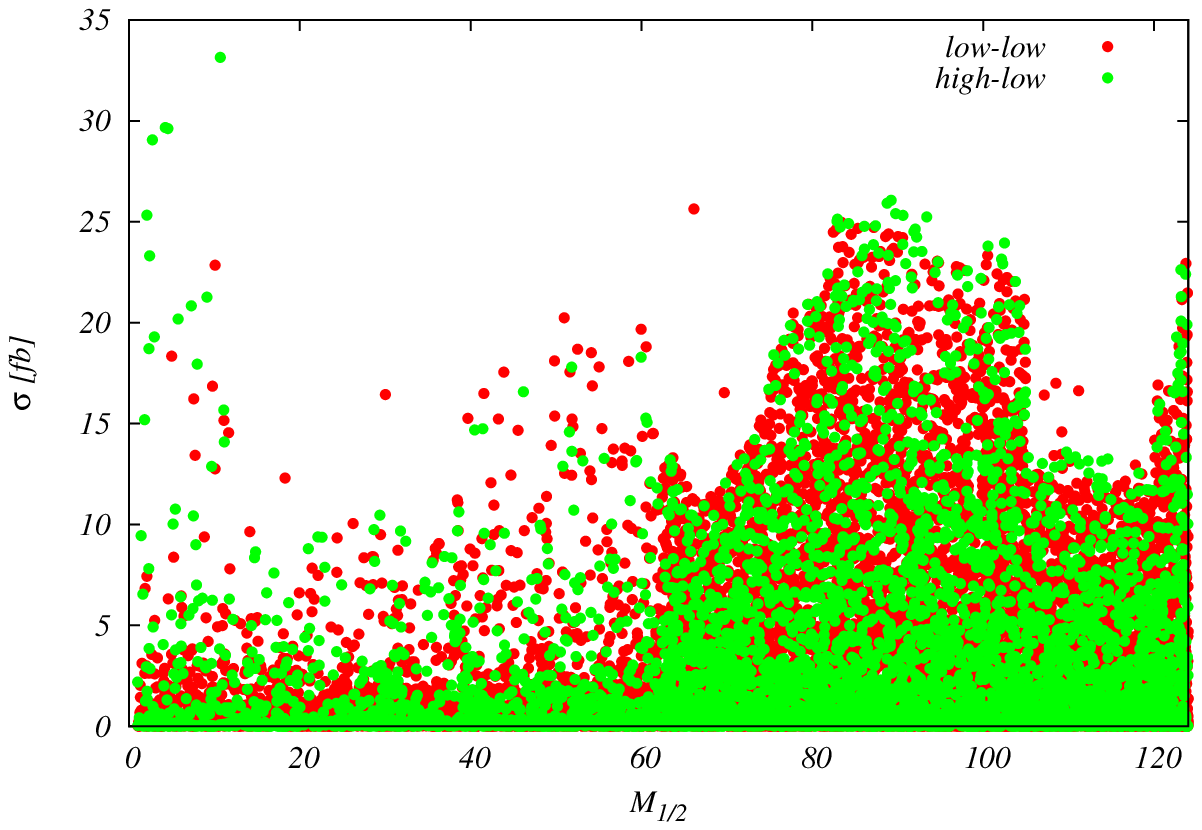}
\includegraphics[width=0.49\textwidth]{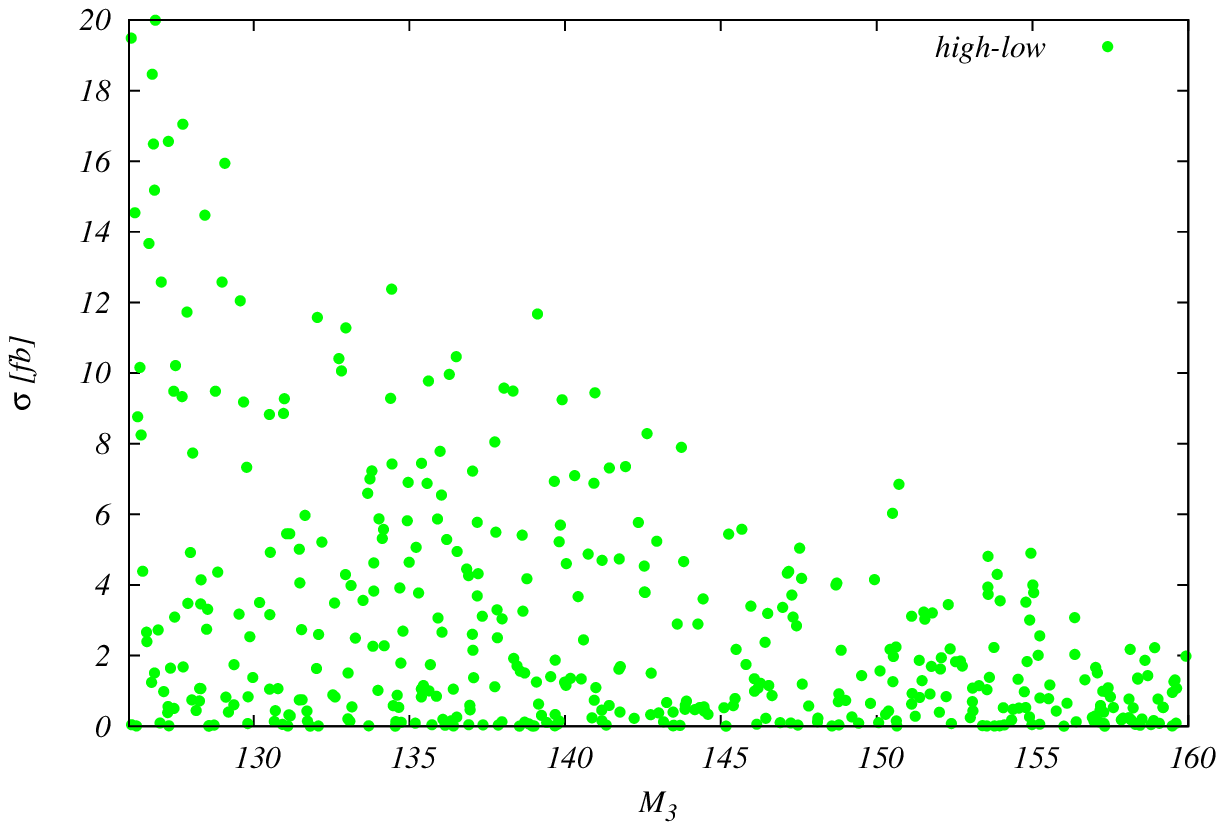}
\caption{\label{fig:xsecs} Maximal production cross section for Higgs-Strahlung for scalars of masses $\neq\,125\,\GeV$ in the TRSM for points passing all discussed constraints. Production cross sections depend on the parameter point and can reach up to 30 \fb. Taken from \cite{Robens:2022zgk}.}
\end{center}
\end{figure}
\end{center}

Here, $|\sin\al|$ is symbolic for the respective mixing angle, with $\sin\al\,=\,0$ denoting complete decoupling. We see that in general, for low mass scalars, mixing angles up to $\sim\,0.3$ are still allowed. 
At Higgs factories, it was shown in \cite{Robens:2022zgk} that $Z\,h$ production is dominant in the low mass range, and also gives the largest contribution to the $\nu\bar{\nu}h$ final state, so we concentrate on Higgs-strahlung. Maximally allowed production cross sections at an $e^+e^-$ collider with a com energy of 250 \GeV~ are displayed in figure \ref{fig:xsecs}.

\begin{center}
\begin{figure}
\begin{center}
\includegraphics[width=0.5\textwidth]{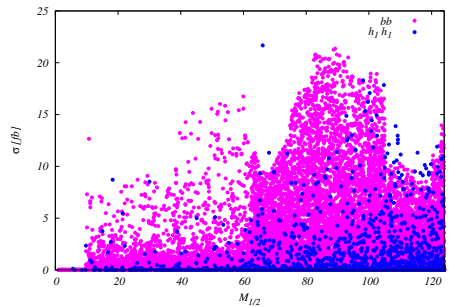}
\caption{\label{fig:bbfin} Production cross sections for $e^+e^-\,\rightarrow\,Z\,h_{1/2}\,\rightarrow\,Z\,X\,X$, with $X\,\equiv\,b$ {\sl (magenta)} and $h_1$ {\sl (blue)}. Points from all data sets are included. Cross sections can reach up to 20 \fb. In the low mass region, also $X\,\equiv\,\tau,c$ final states can become important (not shown here). Taken from \cite{Robens:2022zgk}.}
\end{center}
\end{figure}
\end{center}

As branching ratios for the low mass scalars are inherited via mixing with the scalar from the SM-like doublet, the largest production cross sections are obtained for scenarios where the light scalars decay into $b\bar{b}$ final states; these could in principle be investigated or constrained using already existing projecting bounds at Higgs factories, see discussion in \cite{Robens:2022zgk}. We display cross sections for such final states in figure \ref{fig:bbfin}, together with predictions for $h_1\,h_1$ final states in case $h_2\,\rightarrow\,h_1\,h_1$. For $M_i\,\lesssim\,12\,\GeV$, other final states as e.g. $\tau\,\tau$ and $c\bar{c}$ can lead to cross sections up to $20\,\fb$.

\begin{center}
\begin{figure}
\begin{center}
\includegraphics[width=0.5\textwidth]{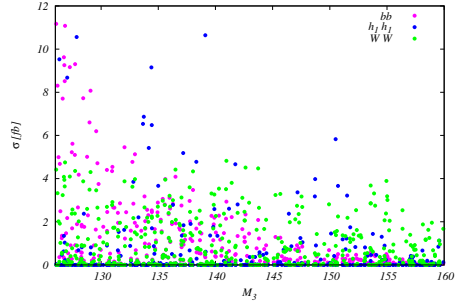}
\caption{\label{fig:finsh} Production cross sections for $e^+e^-\,\rightarrow\,Z\,h_{3}\,\rightarrow\,Z\,X\,X$, with $X\,\equiv\,b$ {\sl (magenta)}, $h_1$ {\sl (blue)}, and $W$ {\sl (green)}. Points from all data sets are included. Cross sections can reach up to $\sim\,12\,\fb$. Taken from \cite{Robens:2022zgk}.}
\end{center}
\end{figure}
\end{center}

For scenarios with $M_3\,\gtrsim\,126\,\GeV$, $h_1\,h_1,\,W^+\,W^-,$ and $b\bar{b}$ can dominate, depending on the specific parameter point. We display the corresponding production cross sections in figure \ref{fig:finsh}. 

\begin{center}
\begin{figure}
\begin{center}
\includegraphics[width=0.5\textwidth]{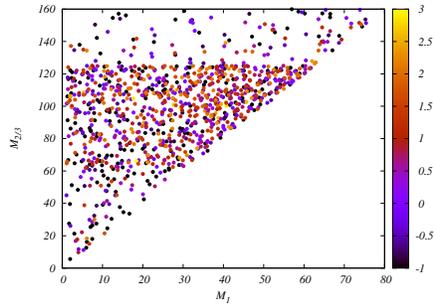}
\caption{\label{fig:Hhh} Production cross sections for $e^+e^-\,\rightarrow\,Z\,h_{2/3}\,\rightarrow\,Z\,X\,X$, with $X\,\equiv\,h_1$, in the $\lb M_1,\,M_{2/3} \rb$ plane. Color coding refers to the $\log_{10}\left[\sigma/\fb\right]$ for $Z h_1 h_1$ production. Maximal cross sections are around 20 \fb. Taken from \cite{Robens:2022zgk}.}
\end{center}
\end{figure}
\end{center}

Finally, I display cross sections $e^+e^-\,\rightarrow\,Z\, h_{2/3}$, with subsequent decays to $h_1\,h_1$ final states in figure \ref{fig:Hhh}. We find the largest cross section of about $\sim\,20\,\fb$ for a parameter point where $M_2\,\sim\,66\,\GeV,\,M_1\,\sim\,18\,\GeV$. The $h_1$ in this parameter point decays predominantly into $b\,\bar{b}$ final states with a branching ratio of about $85\%$.

\subsection{Experimental searches with TRSM interpretations}
\begin{center}
\begin{figure} 
\includegraphics[width=\textwidth]{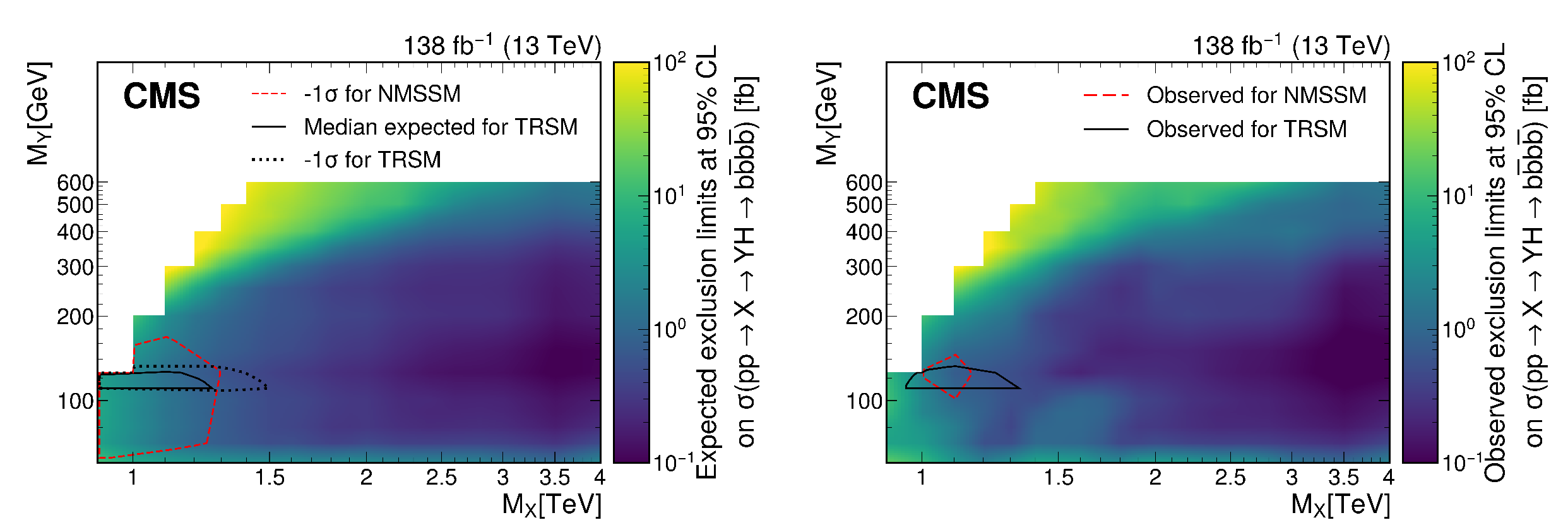}
\caption{\label{fig:cmsres} Expected {\sl (left)} and observed {\sl (right)} $95\%$ confidence limits for the $p\,p\,\rightarrow\,h_3\,\rightarrow\,h_2\,h_1$ search, with subsequent decays into $b\bar{b}b\bar{b}$. For both models, maximal mass regions up to $m_3\,\sim\,\,1.4\TeV,\;m_2\,\sim\,\,140\,\GeV$ can be excluded. Figure taken from \cite{CMS:2022suh}.}
\end{figure}
\end{center}

\begin{center}
\begin{figure} 
\begin{center}
\begin{minipage}{0.45\textwidth}
\begin{center}
\includegraphics[width=\textwidth]{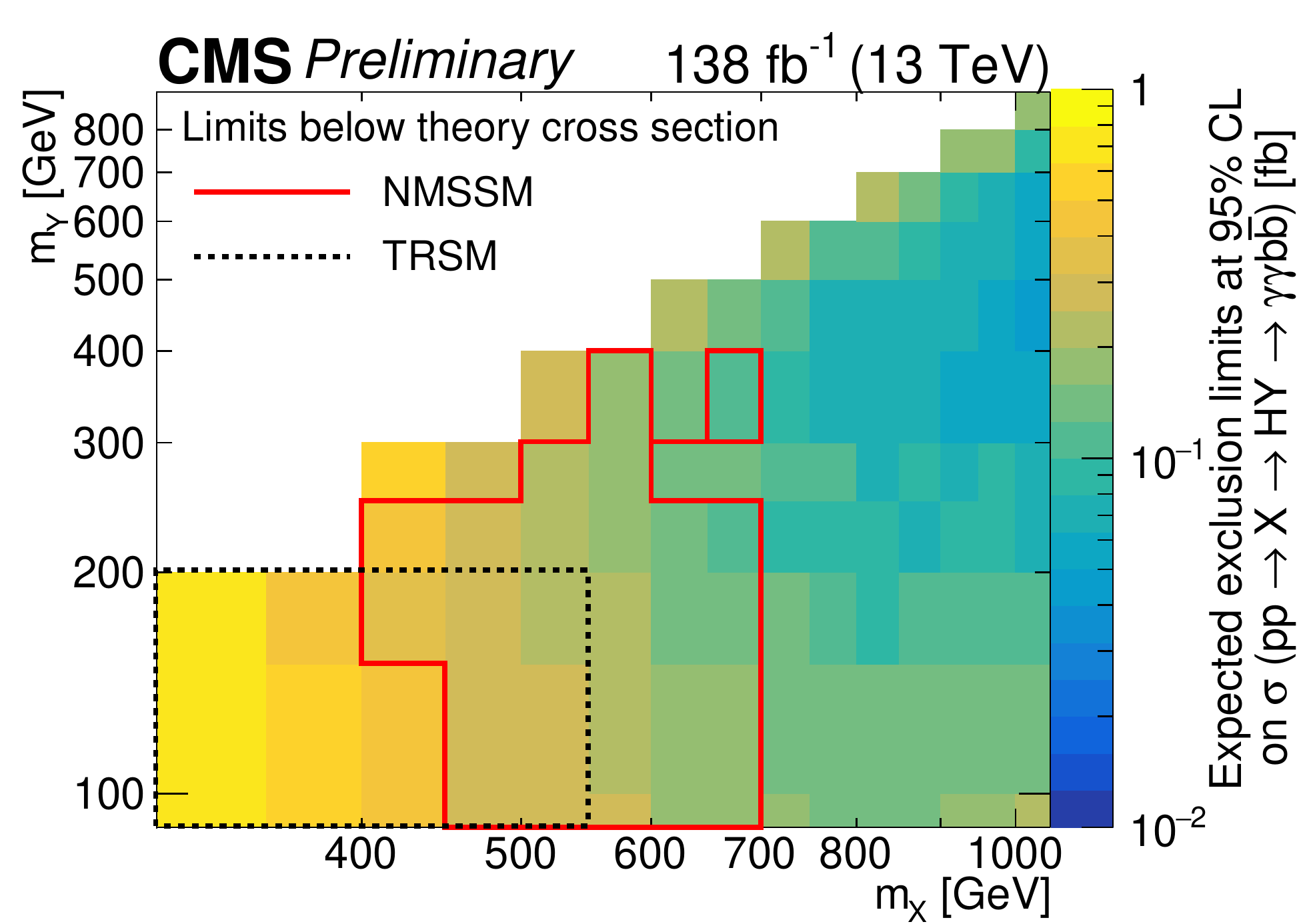}
\end{center}
\end{minipage}
\begin{minipage}{0.45\textwidth}
\begin{center}
\includegraphics[width=\textwidth]{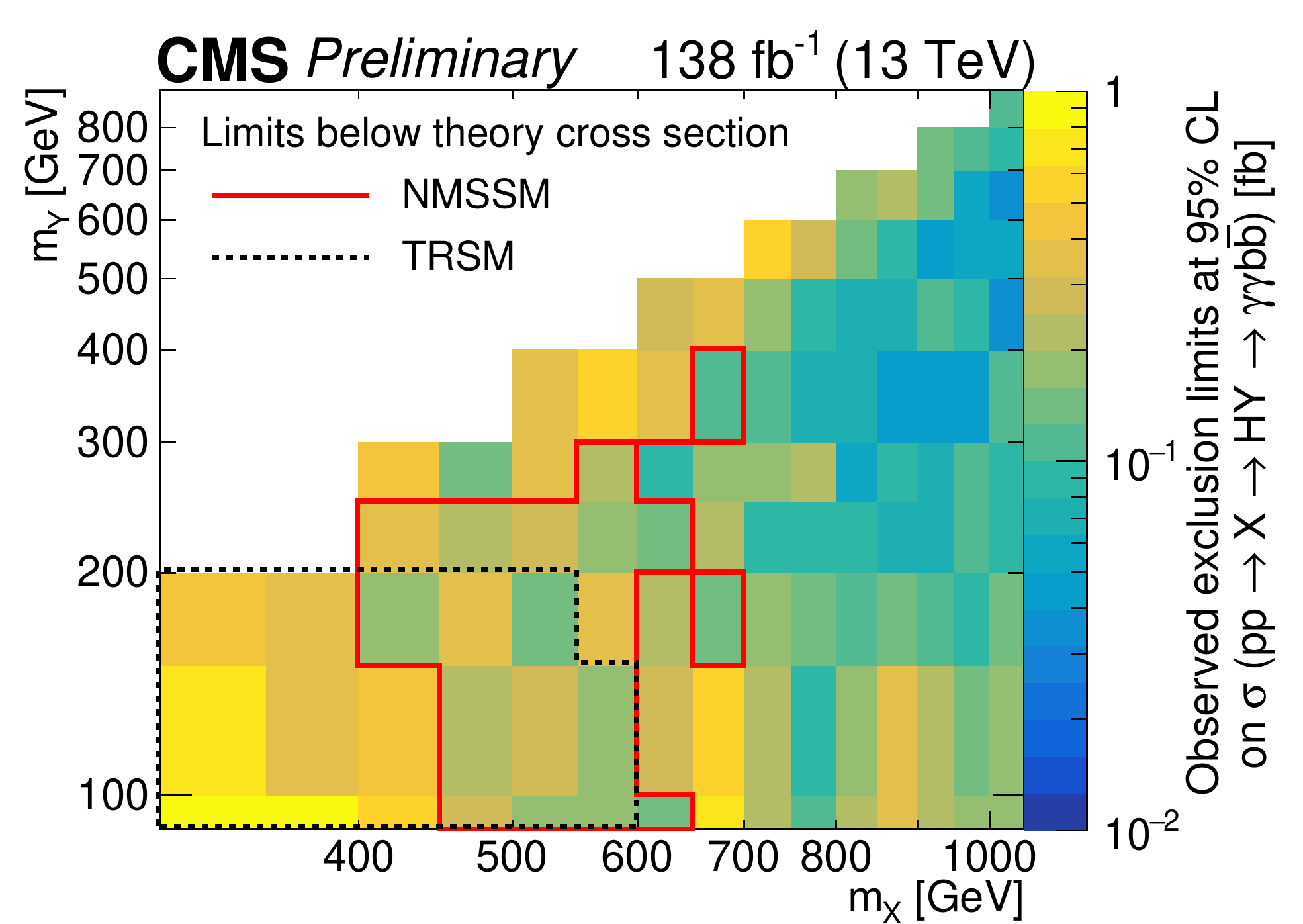}
\end{center}
\end{minipage}
\caption{\label{fig:cmsbbgaga} Expected {\sl (left)} and observed {\sl (right)} $95\%$ confidence limits for the $p\,p\,\rightarrow\,h_3\,\rightarrow\,h_2\,h_1$ search, with subsequent decays into $b\bar{b}\gamma\gamma$. Depending on the model, maximal mass regions up to $m_3\,\sim\,\,800\GeV,\;m_2\,\sim\,\,400\,\GeV$ can be excluded. Figure taken from \cite{CMS-PAS-HIG-21-011}.}
\end{center}
\end{figure}
\end{center}

I also want to comment on experimental searches at the LHC that have made use of TRSM interpretations. Both were performed by the CMS experiment and consider asymmetric production $H_3\,\rightarrow\,h_1\,h_2$, with subsequent decay into $b\bar{b}b\bar{b}$  final states \cite{CMS:2022suh}, as well as $b\bar{b}\gamma\gamma$ in \cite{CMS-PAS-HIG-21-011}. For this, maximal production cross sections were provided in the parameter space, allowing all additional new physics parameter to float; the respective values have been tabulated in \cite{reptr,trsmbbgaga}. In figures \ref{fig:cmsres} and \ref{fig:cmsbbgaga}, the expected and observed limits in these searches are displayed for the TRSM and NMSSM \cite{Ellwanger:2022jtd}.

Several additional searches also investigate decay chains that can in principle be realized within the TRSM, as e.g. other searches for the same final states \cite{CMS:2022qww} or $b\,\bar{b}\mu^+\mu^-$ \cite{ATLAS:2021hbr} final states.

\section{Summary and Outlook}

In these proceedings, I discussed novel results which build on the models presented in \cite{Robens:2022zav}, which were in the same form presented at the Capri Workshop. In particular, I focussed on scenarios with singlet extensions in the SM scalar sector. For a simple real singlet extension, I showed updated results in the $\lb M_H;|\sin\al|\rb$ plane as well as a comparison of possible maximal rates with current bounds for di-Higgs production. I also commented on the possibility to constrain regions in the models parameter space that allow for a strong first-order electroweak phase transition via future precision measurements in the electroweak sector.

The second model I presented is the TRSM, where the scalar sector is augmented by two additional scalars that obey a $\mathbb{Z}_2\,\times\,\mathbb{Z}_2'$ symmetry. Here, I showed predictions for light scalar production at future Higgs factories, as well as some sample results of current LHC searches interpreted within this model.

In summary, the discussion and investigation of BSM scenarios is ongoing. The models discussed here are not subject to flavour constraints, but mainly testable via precision measurements as well as direct searches for resonances. For all scenarios, the continuation of a strong experimental program at colliders is indispensable.

\section*{Acknowledgement}
I thank the organizers of the workshop for additional financial support, as well as A. Papaefstathiou and G. White for fruitful collaboration.


\end{document}